\documentclass[11pt,a4paper]{article}
\pdfoutput=1

\usepackage{jheppub}
\usepackage{rotating}

\usepackage{wrapfig}

\usepackage{amsfonts}
\usepackage{amsmath}
\usepackage{slashed}
\usepackage{amssymb}
\usepackage{latexsym}
\usepackage{multirow}
\usepackage{hyperref}
\usepackage{enumitem}
\hypersetup{colorlinks=true,citecolor=red,linkcolor=magenta,urlcolor=blue}
\usepackage[caption=false]{subfig}
\usepackage{relsize}
\usepackage{booktabs}
\usepackage{cleveref}
\usepackage{fancyhdr}
\usepackage{float}
\usepackage{graphicx}
\usepackage{microtype}
\usepackage{tabularx}
\usepackage{xcolor}
\usepackage[bottom]{footmisc}
\usepackage{shorthand}

\newenvironment{custdescription}{%
   
   \begin{description}[leftmargin=0.25cm, style=sameline]%
}{%
   \end{description}%
}

\usepackage{physics}

\newcommand{\bdash}{\multicolumn{1}{c}{\textendash}}

\newcolumntype{C}{>{$}c<{$}}
\newcolumntype{L}{>{$}l<{$}}
\newcolumntype{R}{>{$}r<{$}}

\newcommand{\diff}[1]{\ensuremath{\frac{1}{\sigma} \frac{d\sigma}{d #1}}}
\newcommand{\ddiff}[2]{\ensuremath{\frac{1}{\sigma} \frac{d\sigma}{d #1 d #2}}}

\title{Improved Constraints on Effective Top Quark Interactions using Edge Convolution Networks}
\author[]{Oliver Atkinson,}  
\author[]{Akanksha Bhardwaj,}
\author[]{Stephen Brown,}
\author[]{Christoph Englert,} 
\author[]{David~J.~Miller,}
\author[]{and Panagiotis Stylianou}
\affiliation[]{School of Physics \& Astronomy, University of Glasgow, Glasgow G12 8QQ, United Kingdom}
\emailAdd{o.atkinson.1@research.gla.ac.uk} 
\emailAdd{akanksha.bhardwaj@glasgow.ac.uk}
\emailAdd{s.brown.7@research.gla.ac.uk}
\emailAdd{christoph.englert@glasgow.ac.uk}
\emailAdd{david.j.miller@glasgow.ac.uk}
\emailAdd{p.stylianou.1@research.gla.ac.uk}
\abstract{We explore the potential of Graph Neural Networks (GNNs) to improve the performance of high-dimensional effective field theory parameter fits to collider data beyond traditional rectangular cut-based differential distribution analyses. In this study, we focus on a SMEFT analysis of $pp \to t\bar t$ production, including top decays, where the linear effective field deformation is parametrised by thirteen independent Wilson coefficients. The application of GNNs allows us to condense the multidimensional phase space information available for the discrimination of BSM effects from the SM expectation by considering all available final state correlations directly. The number of contributing new physics couplings very quickly leads to statistical limitations when the GNN output is directly employed as an EFT discrimination tool. However, a selection based on minimising the SM contribution enhances the fit's sensitivity when reflected as a (non-rectangular) selection on the inclusive data samples that are typically employed when looking for non-resonant deviations from the SM by means of differential distributions.}

\begin{document}
\maketitle
\section{Introduction}
\label{sec:intro}
The search for new physics, albeit so far unsuccessful at the Large Hadron Collider (LHC) remains a cornerstone of
the collider phenomenology programme. The lack of direct evidence for new states beyond 
the Standard Model (BSM) can be interpreted as an indication of a separation between the electroweak scale and the scale
of new physics $\Lambda$. Integrating out BSM states can then be cast into a consistent and systematic extension
of the SM by higher dimensional operators ${\cal{O}}$ with dimensions $[{\cal{O}}]>4$~\cite{Weinberg:1978kz}. Turning the argument around, constraints on any extension
of the SM can be obtained by agnostically reflecting all a priori allowed operator deformations in the interpretation of LHC
results up to a considered order in the $\Lambda^{-1}$ expansion. At operator level up to dimension six in Standard Model Effective Theory (SMEFT)~\cite{Buchmuller:1985jz,Burges:1983zg,Leung:1984ni,Hagiwara:1986vm,Grzadkowski:2010es,Dedes:2017zog}
\begin{equation}
\label{eq:eft}
{\cal{L}}={\cal{L}}_{\text{SM}} + \sum_i {C_i\over \Lambda^2} {\cal{O}}_i\,,
\end{equation}
this programme has been comprehensively addressed from a theoretical perspective (see e.g.~\cite{Alonso:2013hga,Jenkins:2013wua,Jenkins:2013zja,Baglio:2019uty,Corbett:2021eux,Dawson:2021ofa,Dawson:2022bxd} and~\cite{Brivio:2017vri} for a recent review) and provides
a theoretically consistent interpretation at this order in the $\Lambda^{-1}$ expansion. Related phenomenological (proof-of-principle) analyses have focussed on (combinations of) Higgs physics~\cite{Corbett:2015ksa,Corbett:2015mqf,Englert:2015hrx,Englert:2017aqb,Ellis:2014jta,deBlas:2019rxi,Dawson:2018dcd,Brivio:2019myy}, electroweak precision observables~\cite{Ellis:2018gqa,DeBlas:2019qco,Ellis:2020unq}, and the top sector~\cite{Buckley:2015nca,Buckley:2015lku,Vos:2016til,Castro:2016jjv,Aguilar-Saavedra:2018ksv,Hartland:2019bjb,Brivio:2019ius,Ethier:2021bye}, owing to good statistical and systematic control at past and present collider experiments, and their expected roles in BSM physics in general.
 
The experimental implementation of such a strategy is far from trivial. The number of involved and independent effective interactions can be large, thus potentially limiting the sensitivity of a single, specific analysis. In parallel, systematic uncertainties can lead to weak constraints, giving only loose and perhaps non-perturbative limits when understood as UV constraints in concrete matching calculations. 

There are two avenues to improve phenomenological sensitivity. Firstly, decreasing the theoretical and experimental uncertainties alongside systematics of the EFT and SM hypotheses, possibly via data-driven approaches, will lead to improved limits when more data become available (assuming agreement with the SM prevails). Lower limits on the direct evidence of new states, e.g., via $s$-channel production are predominantly driven by the available LHC centre-of-mass energy. Therefore, the lower limit on $\Lambda$ in~\cref{eq:eft}, which is driven by the LHC's energy coverage, will not change dramatically in the future. Thus, any modelling improvement at scales $|Q^2|\ll \Lambda^2$ where the EFT expansion can be considered reliable will be reflected in improved constraints on the Wilson coefficients (WCs) $C_i$ (modulo remaining blind directions).

Secondly, we can resolve to a more comprehensive extraction of information from experimental data. Such strategies are highlighted in the recent resurgence of machine learning (ML) applications to particle physics~\cite{Brehmer:2017lrt,Brehmer:2016nyr,Brehmer:2018kdj,Brehmer:2018eca,DHondt:2018cww,Chatterjee:2021nms,Araz:2021wqm,Konar:2021zdg,Karagiorgi:2021ngt,Bauer:2021gup,Feickert:2021ajf} (in particular focusing also on experimental improvements~\cite{Moreno:2019bmu,Pata:2021oez}). `Traditional' collider observables such as transverse momenta, angles and (pseudo)rapidities, alongside rectangular cuts on these, might not fully capture the exclusion potential when all ad hoc modifications of correlations are considered, which is the key motivation of the EFT approach (in particular this extends to the inclusion of systematic uncertainties~\cite{Englert:2018cfo}).

The latter of these two avenues is the focus of this work. We focus on EFT parameter constraints for the top sector~\cite{Buckley:2015nca,Buckley:2015lku,Vos:2016til,Castro:2016jjv,Aguilar-Saavedra:2018ksv,Hartland:2019bjb,Brivio:2019ius,Ethier:2021bye}. In particular, we focus on $pp\to t\bar t$ production with semi-leptonic top decays as this provides a relatively clean channel with good statistical control to discuss ML-improved EFT strategies (see also Ref.~\cite{DHondt:2018cww}). To reflect expected correlations between the final state (i.e. fully showered and hadronised) objects, we employ Graph Neural Networks (GNNs) with Edge Convolution~\cite{zhou2018graph,9046288,gilmer2017neural,wang2019dynamic}. This setup exploits the structure of data as well as the correlations (`edges') of different intermediate and final state particles (`nodes') and is therefore well motivated for particle physics applications~\cite{Dreyer:2020brq,Blance:2021gcs,Atkinson:2021nlt,Qu:2019gqs,Dorigo:2021iyy,Mikuni:2020wpr,Knapp:2020dde,Mikuni:2020qds,Dezoort:2021kfk,Abdughani:2018wrw,Duarte:2020ngm,Ju:2020tbo}.

This paper is organised as follows. In~\cref{sec:eft}, we review the EFT operators relevant for this case study. We also detail our simulation, analysis and fit setup of $t \bar t$ production. Section~\ref{sec:graphs} is devoted to the machine learning aspects of this study: we briefly outline our baseline cuts (taking the experimental analysis of \cite{CMS:2016oae} as guidance), review our ML setup, and discuss input parameters, training and classification. We highlight the performance improvements of a ML-informed top sector fit in~\cref{sec:gnnimpro} and conclude in~\cref{sec:conc}.

\section{Effective interactions for top pair production with leptonic decays}
\label{sec:eft}
Any differential cross section that follows from~\cref{eq:eft} can be written as
\begin{equation}
\label{eq:xsec}
	\hbox{d}\sigma= \hbox{d}\sigma_{\text{SM}}+\frac{C_i}{\Lambda^2} \hbox{d}\sigma_i^{(1)}+ \frac{C_i C_j}{\Lambda^4} \hbox{d}\sigma_{ij}^{(2)},
\end{equation}
where the ${C_{i}}$ are the Wilson Coefficients (WCs) and $\Lambda$ is the generic scale of new physics (NP). The first term shows the contribution from the Standard Model (SM) only, while the second term is the contributions from the interference of the EFT and the SM terms. The  third term represents the contribution from the EFT squared or cross-terms which are $\Lambda^4$ suppressed. In the following, we will limit ourselves to dimension 6 (differential) cross sections $\sim \Lambda^{-2}$ that result from interference of the EFT and SM amplitudes. While this is a theoretically consistent approach, it also constitutes a conservative case for EFT limit setting: contributions $\sim \Lambda^{-4}$ typically show a dramatic momentum-transfer enhanced behaviour and are therefore relatively easy to constrain, even using standard approaches. Put differently, any sensitivity improvement that we can identify for the linearised approach will generalise to the inclusion of the $\sim \Lambda^{-4}$ terms in~\cref{eq:xsec}.

\subsection{Analysis Setup and Fit Methodology}
\label{sec:details}

\begin{table}[!t]
  \centering
  \begin{tabular}{ccR}
    \toprule
    Distribution                           & Observable         & \multicolumn{1}{c}{Binning}                        \\
    \midrule
    \diff{|y_t^h|}                         & $|y_t^h|$          & [0.0, 0.2, 0.4, 0.7, 1.0, 1.3, 1.6, 2.5]        \\
    \diff{|y_t^l|}                         & $|y_t^l|$          & [0.0, 0.2, 0.4, 0.7, 1.0, 1.3, 1.6, 2.5]        \\
    \diff{|y_{t\bar{t}}|}                  & $|y_{t\bar{t}}|$   & [0.0, 0.2, 0.4, 0.6, 0.9, 1.3, 2.3]             \\
    \diff{p_{\perp}^{t,h}}                 & $p_{\perp}^{t,h}$  & [0, 45, 90, 135, 180, 225, 270, 315, 400, 800]~\text{GeV}  \\
    \diff{p_{\perp}^{t,l}}                 & $p_{\perp}^{t,l}$  & [0, 45, 90, 135, 180, 225, 270, 315, 400, 800]~\text{GeV}  \\
    \diff{m_{t\bar{t}}}                    & $m_{t\bar{t}}$     & [300, 375, 450, 530, 625, 740, 850, 1100, 2000]~\text{GeV} \\
    \ddiff{|y_{t\bar{t}}|}{|m_{t\bar{t}}|} & $|y_{t\bar{t}}|$   & [0.0, 0.2, 0.4, 0.6, 0.9, 1.3, 2.3] \\
                                  & $m_{t\bar{t}}$     & [300, 375, 450, 625, 850, 2000]~\text{GeV}                \\
    \ddiff{p_{\perp}^{t, h}}{|y_t^h|}      & $p_{\perp}^{t, h}$ & [0, 45, 90, 135, 180, 225, 270, 315, 400, 800]~\text{GeV}\\
                                        & $|y_t^h|$          & [0.0, 0.5, 1.0, 1.5, 2.5]                       \\
    \bottomrule
  \end{tabular}
  \caption{Distributions provided in Ref.~\cite{CMS:2016oae} and included in the fit in this work.}
  \label{tab:data}
\end{table}

We use the {\tt{SMEFTSim}}~\cite{Brivio:2017btx,Brivio:2020onw} implementation to include the effective operators, which is then interfaced with {\tt MadGraph5}~\cite{Alwall:2014hca} via {\tt{FeynRules}}~\cite{Alloul:2013bka} and {\tt{UFO}}~\cite{Degrande:2011ua} to generate the event samples at leading order (LO)\footnote{In this work, we focus on GNN performance of EFT parameter fits and limit ourselves to a leading order analysis. We note that including higher order contributions for the SM hypothesis is crucial to obtain consistency with the measured data, but will not impact the qualitative results of this work. We have checked that the results of~\cref{tab:baseline_bounds} are qualitatively reproduced by a full NLO fit using a forthcoming version of {\sc{TopFitter}}~\cite{toapp}.} for
\begin{equation}
\label{eq:process}
p p \to t\bar t \to \ell  b \bar b j + {\slashed{E}}_T\,.
\end{equation}
We use a $\sqrt{s}=13~\text{TeV}$ analysis by the CMS collaboration~\cite{CMS:2016oae} as inspiration to investigate (correlated) differential measurement results and representative data binning as given in~\cref{tab:data}. SM predictions are injected as mock reference data for the luminosity~${\cal{L}}_{\text{ref}}=2.3~\text{fb}^{-1}$ of Ref.~\cite{CMS:2016oae} and we scale statistical uncertainties relative to this luminosity, using ${\sqrt{{\cal{L}}_{\text{ref}}/{\cal{L}}}}$ for extrapolations. Our implementation relies on {\tt{Rivet}}~\cite{Buckley:2010ar,Bierlich:2019rhm}, which processes events after showering with {\tt{Pythia8}}~\cite{Sjostrand:2014zea} before feeding them into the fit. 

To avoid imposing any assumptions as to correlations --- and remove the chance that double-counting of events would artificially inflate sensitivity to EFT contributions --- a single distribution is used where bin-to-bin correlations are included, and a single bin is used where they are not. In the absence of a full reference correlation/covariance matrix the selection of the bin/distribution is made on a coefficient-by-coefficient basis, with the input with maximum deviation from a fixed point on that axis being selected. This maximum sensitivity, minimum correlation assumption input subset is then used to determine individual and profiled bounds for the coefficient being studied. Where a normalised distribution is used we must drop a bin, as otherwise the covariance matrix will be singular. The dropped bin is chosen such that we obtain the most stable covariance matrix, with the bin with the largest uncertainty being dropped if there are multiple bins leading to an equivalently well-conditioned covariance matrix.\footnote{For details on statistical inference we refer interested readers to Refs.~\cite{Buckley:2015nca,Buckley:2015lku}.}

In the following we will consider bounds for all relevant operators using the dimensionless `bar' notation
\begin{equation}
\bar{C}_i = C_i \, {v^2\over \Lambda^2}\,,
\end{equation}
with the electroweak
expectation value $v\simeq 246~\text{GeV}$. 

In many standard analyses, cut-and-count techniques are often used to restrict the phase space region in such a way that the SM contamination is minimised, and as a result this yields an increased new-physics sensitivity. However rectangular cuts often yield inferior sensitivity compared to the methodically selected regions by means of machine learning classifiers. In our scenario an efficient event-by-event classification using GNNs, separating the generated events into either pure SM or the SMEFT operators that sourced them, could lead to improvements on the bounds of WCs after imposing cuts on the output score of the network.

\begin{figure*}[!b]
	\centering
	\includegraphics[width=0.65\textwidth]{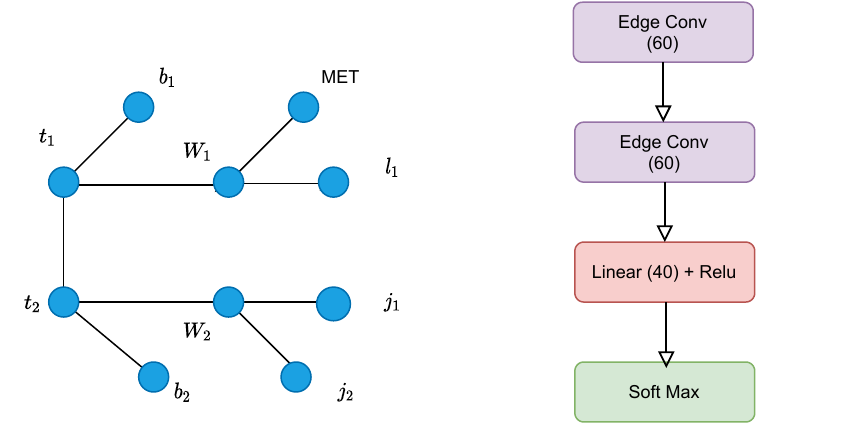}
	\caption{Representative diagram for the input graph and the network architecture used in this paper.}
	\label{fig:Network_arch}
\end{figure*}

\section{Graph representation of events}
\label{sec:graphs}
In order to use a GNN as a classifier, the events need to be embedded in a graph structure with nodes, edges and features associated to observables of final states or reconstructed objects. While various different approaches are possible to construct a graph from the IR-safe, calibratable and detectable final states, we employ a physics-motivated strategy, creating graphs similar to the tree of the chain of~\cref{eq:process} \footnote{We also checked the fully connected graph and found low performance for the given network as the number of edges increases, which carry much less physics information. Hence the decay chain-like structure in the graph gives good performance.}. Concretely, we pre-process the data samples and require at least two jets of transverse momentum $p_T(j) > 20~\text{GeV}$ and pseudorapidity $\abs{\eta} < 5$ that are not b-tagged. The event is vetoed if there are not at least two $b$-jets and one lepton $\ell$ in the central part of the detector ($\abs{\eta(\ell)} < 2.5$), where the b-jets must also satisfy $p_T(b) > 20~\text{GeV}$. Subsequently, we embed the passed events into graphs using the following steps (see also~\cref{fig:Network_arch}):
\begin{custdescription}
\item[(i)~Nodes:] Firstly, the missing transverse momentum (MTM) is identified by balancing the net visible momenta, $-p(\text{visible})$, neglecting the longitudinal components. A node is added corresponding to MTM. Then, for each lepton, we attempt to reconstruct the $W$ four-momentum as a sum of the lepton's four-momentum and the MTM. The invariant mass of the $W$ candidate is calculated and if it falls within $\left[65, 95\right]~ \text{GeV}$ a node is added, labelled $W_1$, as well as one for the b-jet $b_1$ that has the smallest separation $\Delta R = \sqrt{\Delta \eta^2 + \Delta \phi^2}$ from $W_1$. In the case where there are more lepton-MTM combinations with compatible invariant mass, the one closest to the $W$ boson mass is selected. The top from the leptonic decay chain $t_1$ is finally reconstructed from the four-momenta of $\ell$, $b_1$ and MTM and obtains its respective node. Following a similar procedure, we consider combinations of jets to find a pair with dijet invariant mass $70~\text{GeV} \leq m(jj) \leq 90~\text{GeV}$. If a pair is found we add nodes for the two jets $j_1$, $j_2$ and for the second boson $W_2$, otherwise we only add nodes for the two leading jets. From the remaining b-jets a node is added for the leading one, $b_2$, as well as for the second top $t_2$ whose four-momentum is reconstructed using $b_2$, $j_1$ and $j_2$. We scan over the remaining particles and if any are within $\Delta R < 0.8$ of any of the identified or reconstructed objects we add a node that will be connected only to the nearby object.
\item[(ii) Edges:] The connections between the nodes create the adjacency matrix of the graph and the nodes of the final states are connected to the ones of the reconstructed objects from which they are derived. We first connect the MTM and lepton to $W_1$ and subsequently, $W_1$ and $b_1$ are connected to the first top quark node. If a $W_1$ was not created then the aforementioned final states connect directly to $t_1$.\footnote{We expect that this will lead to a further enhancement of sensitivity when the $\Lambda^{-4}$ non-resonant contributions are considered.} Similarly, for the other leg of the decay chain, if $W_2$ was successfully reconstructed, we join its node with the two jets used to reconstruct it, and then $W_2$ and $b_2$ are connected to the top node. The jets are directly connected to the top if there is no node for $W_2$. Any node originating from the remaining final states is connected to the node of the object that satisfied $\Delta R < 0.8$.
\item[(iii) Node features:] After constructing the node and edges, we associate each node with a feature vector $[p_T, \eta, \phi, E, m, \text{PID}]$,  which represent transverse momentum, pseudorapidity, azimuthal angle, energy, mass and particle identification number respectively.
\end{custdescription}

%
\subsection{Graph Neural Network with Edge Convolution}
\label{sec:network}
Convolution networks have seen a range of developments in the past few years. These have created the capability to employ multi-scale localised spatial features. However, Convolutional Neural Networks (CNNs) are limited to work on regular Euclidean-data like images. Recent GNN developments have overcome this limitation through generalising CNNs to operate on graph structured data, facilitating the exploration of non-Euclidean domains of the data~\cite{726791}. This was formalised as Message Passing Neural Networks~(MPNNs) in Ref.~\cite{gilmer2017neural} for supervised learning applications. We briefly describe the general paradigm of the MPNN, which we will generalise later for the edge convolution (EdgeConv) network used in this paper. MPNNs have two main components: a message-passing phase and a graph readout layer. The message passing is defined as a mathematical operation between two nodes $i$ and $j$. We define $x_i^{(l)}$ as the $i$th node's features and $\vec{e}^{\,(l)}_{ij}$ as the edge connecting the nodes $i$ and $j$ at the $l$th time-step, where the vector sign represents the directed graph. A graph can be undirected or directed; we have used bi-directed graphs for this study. During the message-passing phase a message $\vec{m}^{\,(l)}_{ij}$ is calculated between the two nodes by the following operation,
\begin{equation}
	\label{eq:msg_pass}
	\vec{m}^{\,(l)}_{ij}= \vec{M}^{\,(l)}(\vec{x}^{\,(l)}_i,\vec{x}^{\,(l)}_j,\vec{e}^{\,(l)}_{ij})\,.
\end{equation}
The message function can be a linear activation function or a multilayer-perceptron~(MLP), which is shared between the edges and is analogous to convolution operation (here we use a linear activation function for the message function). Once the messages between all connected nodes have been calculated in a layer, each node feature is updated using an aggregation function
\begin{equation}
	\label{eq:agg}
	\vec{x}^{\,(l+1)}_i=\vec{A}(\vec{x}^{\,(l)}_i,\{\vec{m}^{\,(l)}_{ij}|\, j\in\mathcal{N}(i)\})\,,
\end{equation}
where $\mathcal{N}(i)$ are the nodes which are connected to $i$th node and $\vec{A}$ is the permutation invariant function (for instance `max', `sum', or `mean'). The vector $\vec{x}^{\,(l+1)}$ is the input to the next message passing layer. For graph classification, after some message passing operation $L$  we perform a permutation invariant graph readout operation $\Box$ on the final node features $x_i^{(L)}$,  

\begin{equation}
\label{eq:agg_graph}
\vec{X}=\Box(\vec{x}^{(L)}_i|i\in G),
\end{equation} where $G$ denotes the input graph. This gives us fixed length representation of (possibly) variable length graphs, and feeds into a downstream neural network. 

We use an EdgeConv network in this study, which is an ideally suited network for exploiting the edge features from given node features. The edge convolution operation is defined with the following message-passing function
\begin{equation}
	\label{eq:edge_conv}
	\vec{x}_{i}^{\,(l+1)} = \frac{1}{|\mathcal{N}(i)|} \sum_{j \in \mathcal{N}(i)} {\text{\sc{ReLU}}}\left(\Theta(\vec{x}_j^{\,(l)} - \vec{x}_i^{\,(l)}) + \Phi(\vec{x}_i^{\,(l)})\right),
\end{equation}
where aggregation for each node is done using `mean', after which the features of each node are updated. The linear layers $\Theta$ and $\Phi$ take the inputs and map them to identical dimensional spaces. We use $L=2$ and mean graph-readout.

\subsection{Network Architecture and Training}
We use the {\tt Deep Graph Library}~\cite{wang2020deep} and {\tt PyTorch}~\cite{paszke2019pytorch} to construct the graphs and the networks that classify the different EFT signal contributions and the SM `background'. Models with different architectures are trained on data samples that consist of 70000 events for each class, with $80\%$, $10\%$ and $10\%$ used for training, validation and testing respectively. The network models considered, incorporate {{EdgeConv}} layers followed by hidden linear layers and {\sc{ReLU}} is used as the activation function for each layer. Probabilities for each class can be obtained from the output layer by applying the softmax function. We choose the categorical cross-entropy loss function for the multi-class classification problem and use the Adam optimiser with a learning rate of 0.001 to minimise the loss function. The learning rate decays with a factor of $0.1$ if the loss function has not decreased for three consecutive epochs. We train the models for 100 epochs in mini-batches of 100 graphs and an early stopping condition when no loss decrease has occurred for ten epochs. 

By varying the amount of layers and nodes, and training the different models on the data, we find that the configuration of two {{EdgeConv}} layers of 60 nodes and one hidden linear layer of 40 nodes performs particularly well for our scenario. Any event used during training or validation is not considered further in any other part of this work. The loss and accuracy curves for the classifying events have been checked to avoid overtraining. It is worth highlighting that we observe signs of overtraining when we consider deeper networks. The good performance of a relatively shallow network signifies that non-resonant physics is characterisable by relatively few phenomenological properties, which is consistent with the findings of traditional differential EFT fits (see in particular Ref.~\cite{Englert:2017aqb}). This observation will form the baseline of the qualitative discussion of a two-operator example in the next section.

\section{GNN-improved Wilson coefficient constraints}
\label{sec:gnnimpro}
\subsection{A minimal example}
\label{sec:example}
For illustration purposes, we first limit our study to a three-class classification problem. The network output in this example returns the probability of an event belonging to each of the three classes. An event is then assigned to the EFT/SM class with the greatest corresponding probability.
Generalising this to a higher and critical number of WCs will be the focus of~\cref{sec:full}. 
\begin{figure*}[!t]
	\centering
	\parbox{0.52\textwidth}{\includegraphics[width=0.52\textwidth]{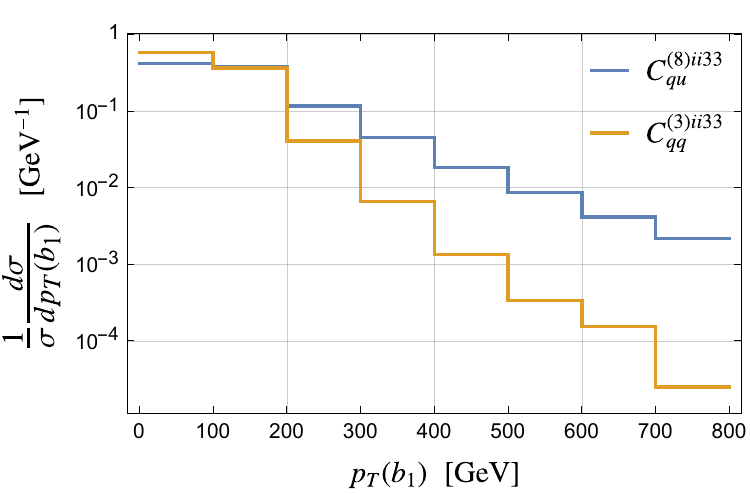}}
	\hspace{0.4cm}
	\parbox{0.4\textwidth}{\vspace{-3.2cm}\caption{\label{fig:ptb1_histos}The normalised $p_T(b_1)$ distributions at the 13 TeV LHC for the two operators of the three-class example, Eq.~\eqref{eq:opstbc}.}}
\end{figure*}

The restriction employed in this section is motivated from the generic modifications that can be expected from EFT interactions. Momentum-dependent interactions will typically enhance the tails of momentum-dependent distributions compared to the SM, 
while interactions that modify SM couplings (feeding into, e.g., a modified top quark width) will predominantly lead to a modified inclusive rate with momentum-related distributions similar to the SM. We reflect this in our choice of operators for this section:
\begin{equation}
\label{eq:opstbc}
	\begin{split}
		{\cal{O}}_{qu}^{(8)ii33} & = (\bar{q_i}\gamma_\mu T^A q_i)(\bar{u_3}\gamma^\mu T^A u_3)\,,\\
		{\cal{O}}_{qq}^{(3)ii33} & = (\bar{q_i}\gamma_\mu \tau^I q_i)(\bar{q_3}\gamma^\mu \tau^I  q_3)\,.
	\end{split}
\end{equation}
The distributions of the hardest $b$ jet for these operators are given in~\cref{fig:ptb1_histos}. Correlated with the events hardness are more central final states and characteristically modified angular and rapidity separations. Identifying the most appropriate superposition of physical observables is therefore critical for a particularly sensitive EFT analysis.
We consider the two operators of~\cref{eq:opstbc} as they exhibit a particularly distinguishable phenomenology, but they will also allow us to discuss the limitations of using different approaches to a ML-informed limit setting.
\begin{figure*}[!t]
	\centering	
	 \includegraphics[height=0.41\textwidth]{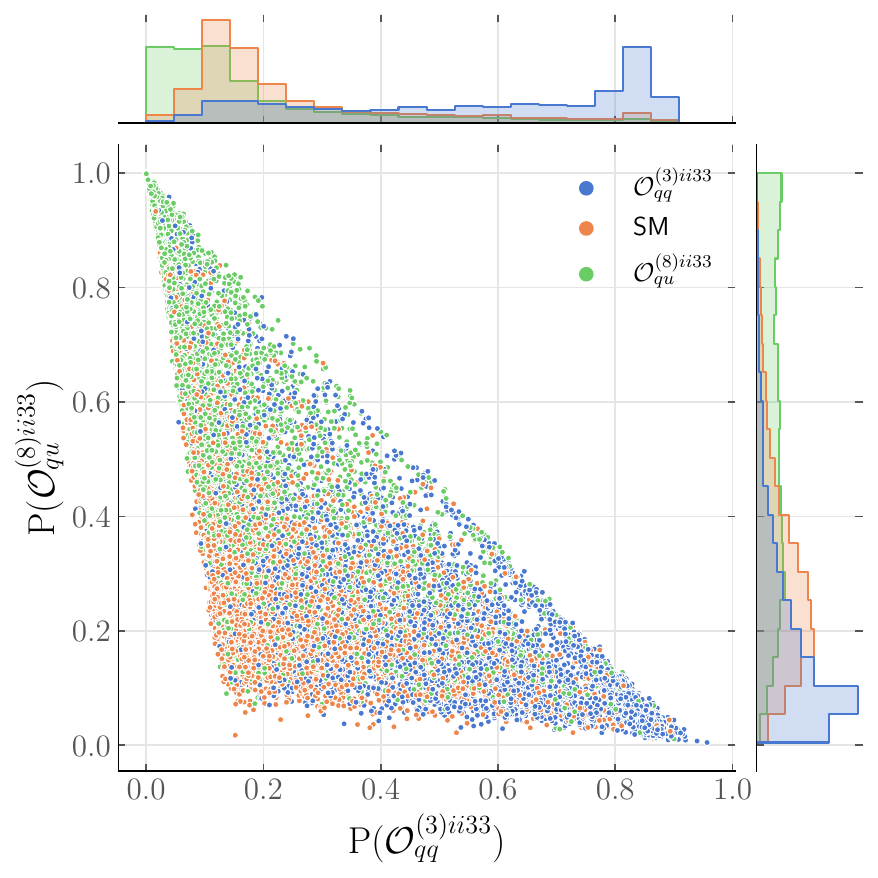}
	\hfill
	\includegraphics[height=0.41\textwidth]{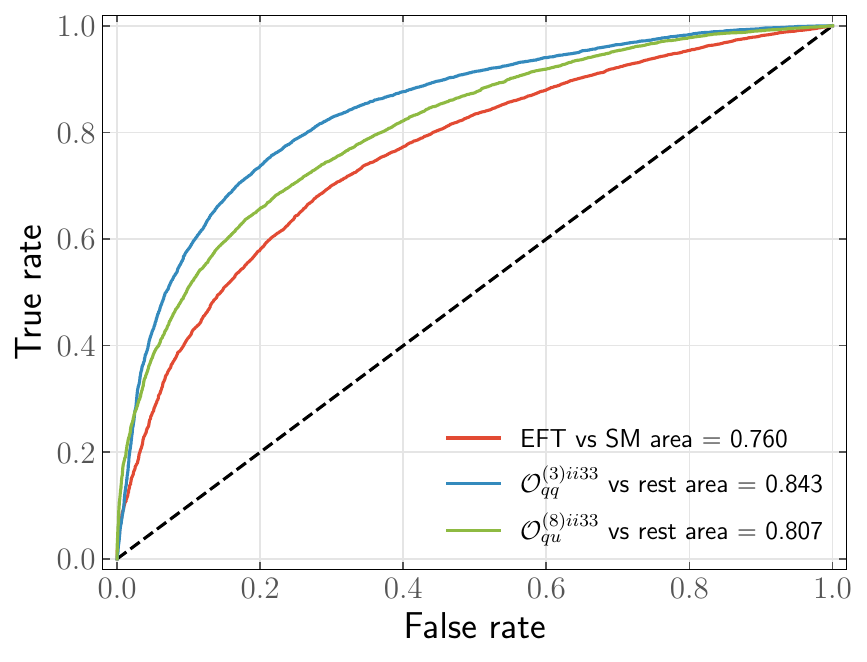}
	\caption{ The probabilities calculated for each event to be a result of each SMEFT insertion is shown. On the right the Receiver Operator Characteristic (ROC) curves are shown. We calculate these in a one-vs-rest scheme for each operator.}
	\label{fig:Network_performance}
\end{figure*}

In~\cref{fig:Network_performance} (left), the probabilities calculated for each event to be a result of each SMEFT insertion are shown. As can be expected, events arising from $\mathcal{O}_{qu}^{8(ii33)}$ are more commonly located in the upper left region (large $\mathcal{O}_{qu}^{8(ii33)}$ probability, small $\mathcal{O}_{qq}^{3(ii33)}$ probability) while events from $\mathcal{O}_{qq}^{3(ii33)}$ are in the bottom right. In contrast, the SM events usually end up in a region where the probabilities for $\mathcal{O}_{qu}^{8(ii33)}$ and $\mathcal{O}_{qq}^{3(ii33)}$ are both low (and the probability of belonging to the SM is high due to the normalisation of probabilities). The network is able to discriminate efficiently among the three classes and different regions can be efficiently removed by cuts on the two output probabilities. On the right in~\cref{fig:Network_performance} the Receiver Operator Characteristic (ROC) curves are shown. We calculate these in a one-vs-rest scheme by first binarising the labels and using the network score output for each WC. We also show an EFT vs SM ROC curve where all EFT labels are marked as signal and the SM as background. We construct the ROC curve using the summed scores for each new physics WC, which we later generalise when more than two contributions are on.

\begin{figure*}[!t]
	\centering
	\parbox{0.46\textwidth}{\includegraphics[height=0.31\textwidth]{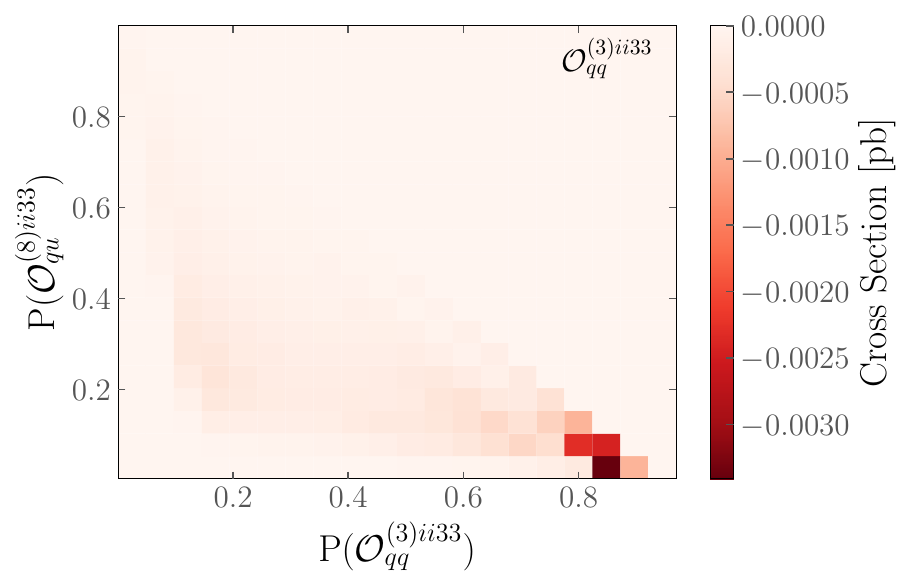}}
	\hfill
	\parbox{0.46\textwidth}{\vspace{0.2cm}\includegraphics[height=0.31\textwidth]{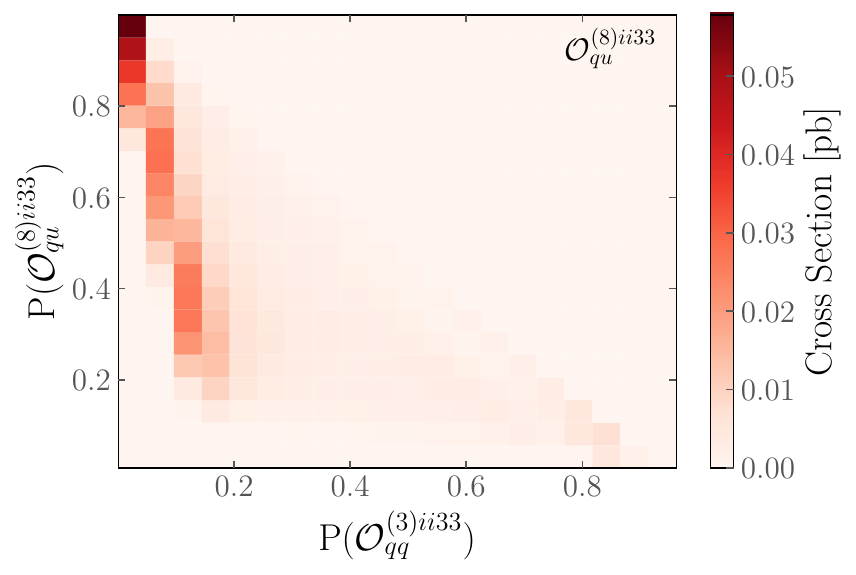}}
	\parbox{0.46\textwidth}{\includegraphics[height=0.31\textwidth]{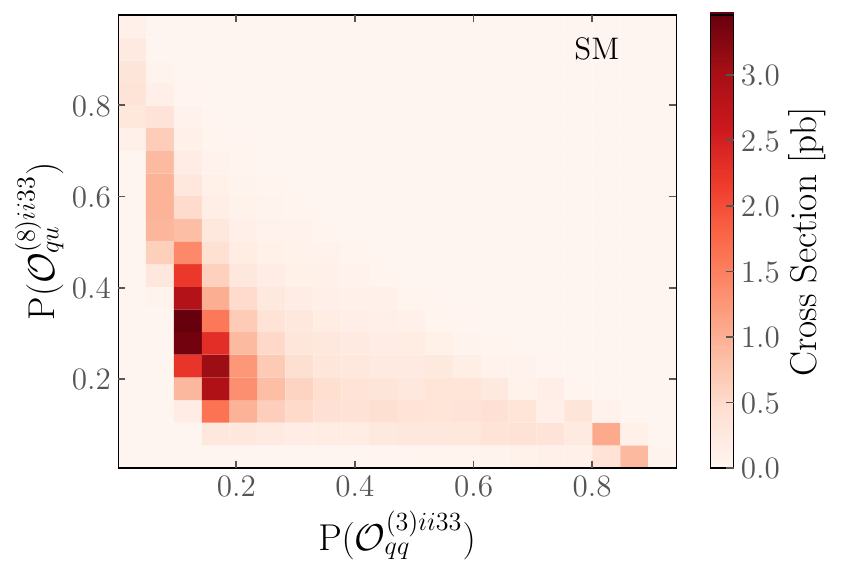}}
	\hfill
	\parbox{0.46\textwidth}{\caption{Example two-dimensional histograms for each contribution, normalised to the cross section rate. \label{fig:2D_hist}}}
\end{figure*}

\begin{figure*}[!b]
	\centering
	\includegraphics[width=0.46\textwidth]{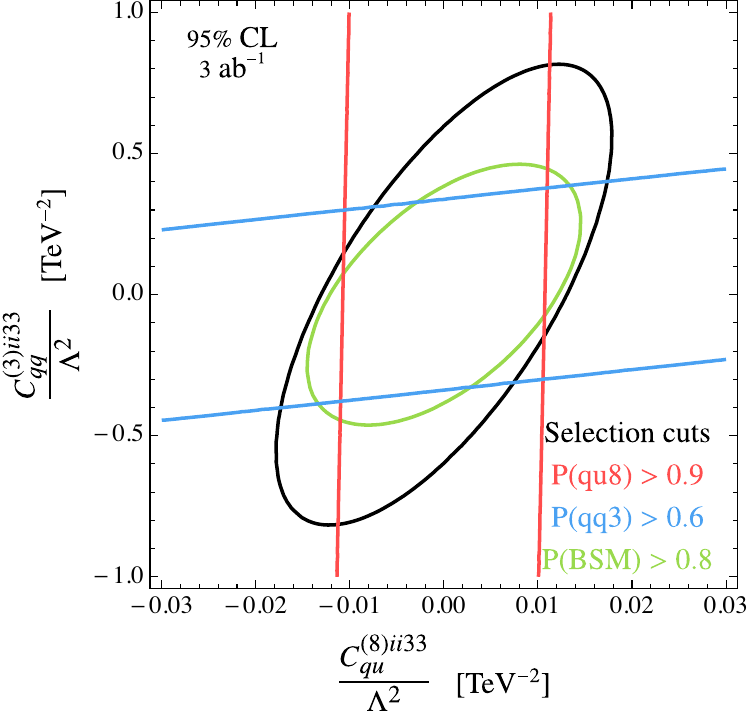}
	\hfill
	\includegraphics[width=0.46\textwidth]{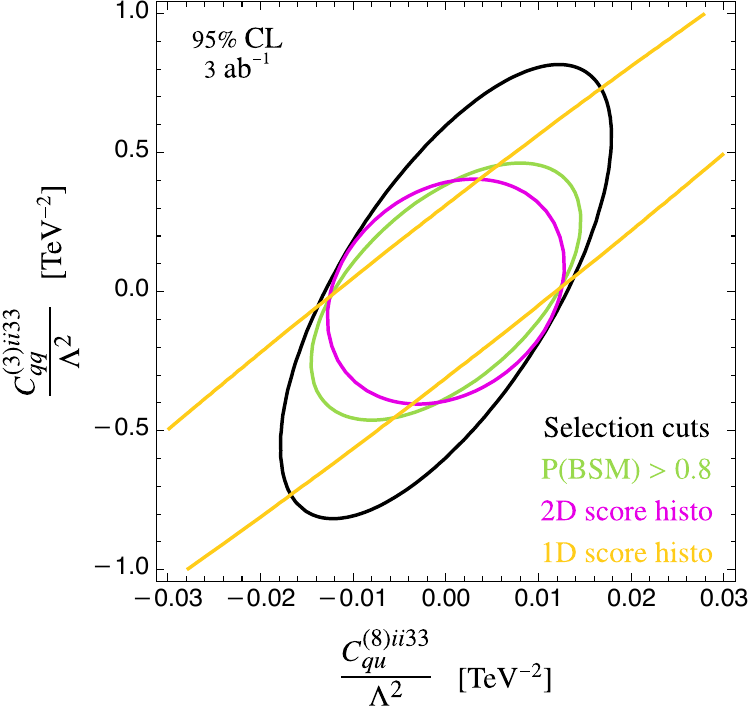}
	\caption{WC constraint contours at the 95\% C.L. from $\chi^2$ fitting; in black from the data of the baseline selection of~\cref{sec:eft} which also passes the network requirements. The left plot shows the contours from cuts on the NN scores at the optimal value of these score cuts, with the analysis performed using $p_T(b_1)$ distributions. The right plot shows the BSM score cut as in the left plot, along with the contour from the 2D score histogram of~\cref{fig:2D_hist} (with no score cuts) analysis, as well as an analysis using the 1D BSM score histogram. For details see text.}
	\label{fig:ML_Contours}
\end{figure*}

To examine the improvement of the network performance for this simplified test case of two WCs modifying SM production, we performed a $\chi^2$ fit for each operator to yield bounds on the WCs. To construct the $\chi^2$ (for details see Ref.~\cite{Buckley:2015lku}), we use the distribution $p_T(b_1)$, the transverse momentum of the leading b-jet. To gain as much statistical control as possible, we also extrapolate the results to an integrated luminosity of 3 ab$^{-1}$, in line with the expected performance of the High-Luminosity (HL) LHC. The qualitative pattern of results, however, is independent of the luminosity chosen. Performing this analysis on the full datasets gives the contours shown in black in~\cref{fig:ML_Contours}, establishing a baseline against which we can evaluate the improvement in the constraints from applying the GNN results.

To demonstrate the power of the GNN approach, we cut on the datasets, based on the probability assigned by the network of belonging to a given class; only events with a probability greater than an optimised value of belonging to one operator class are used in the $\chi^2$ fit. The correlation of~\cref{fig:Network_performance} (left) allows us to select a threshold probability to cut on, which has the effect of substantially reducing the SM background and the contamination from the other operator, resulting in a relatively stronger signal effect and thus a tighter constraint on the WC for the operator for which the cut is performed. This is shown in the blue and red contours in~\cref{fig:ML_Contours}, where the values of the cuts have been tuned to give maximal performance for each operator respectively, whilst avoiding completely depleting bins in the SM $p_T(b_1)$ distribution, as to do so would lead to unrealistic bounds on the WCs as statistical control is lost.

Due to this optimisation the bounds on individual coefficients improve, yet the other coefficient is essentially free, with expectedly far worse performance than in the original case with the full dataset. To resolve this and improve the combined bounds, we consider the probability P(BSM), which is simply the sum of the network assigned probabilities of each operator, i.e.
\begin{equation}
\text{P(BSM)}=\text{P}({\cal{O}}_{qu}^{(8)ii33}) + \text{P}({\cal{O}}_{qq}^{(3)ii33} )
\end{equation}
for the two operator classification considered here. This does indeed result in a combined bound that is superior to the original analysis.

\begin{figure*}[!t]
	\centering
	\includegraphics[width=0.46\textwidth]{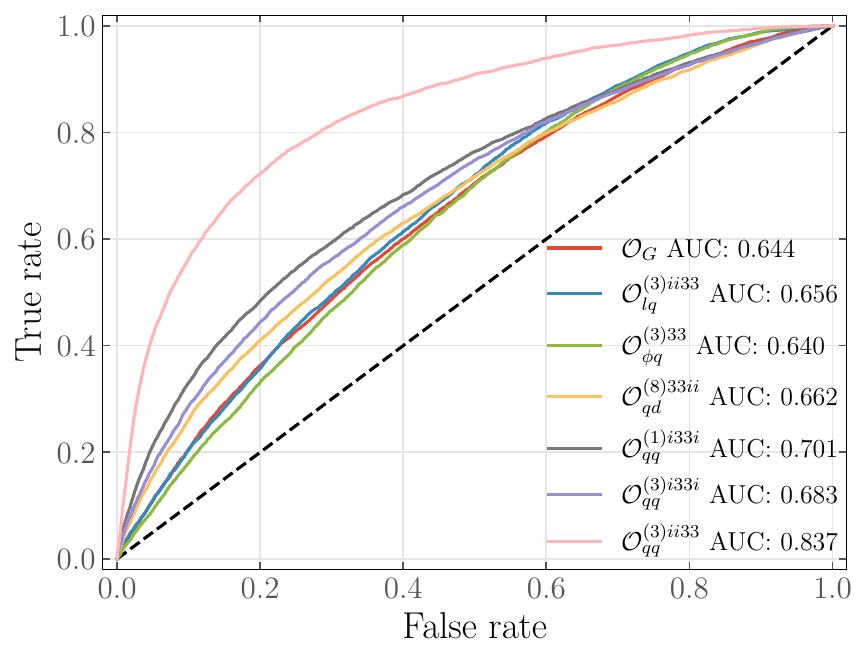}
	\hfill
	\includegraphics[width=0.46\textwidth]{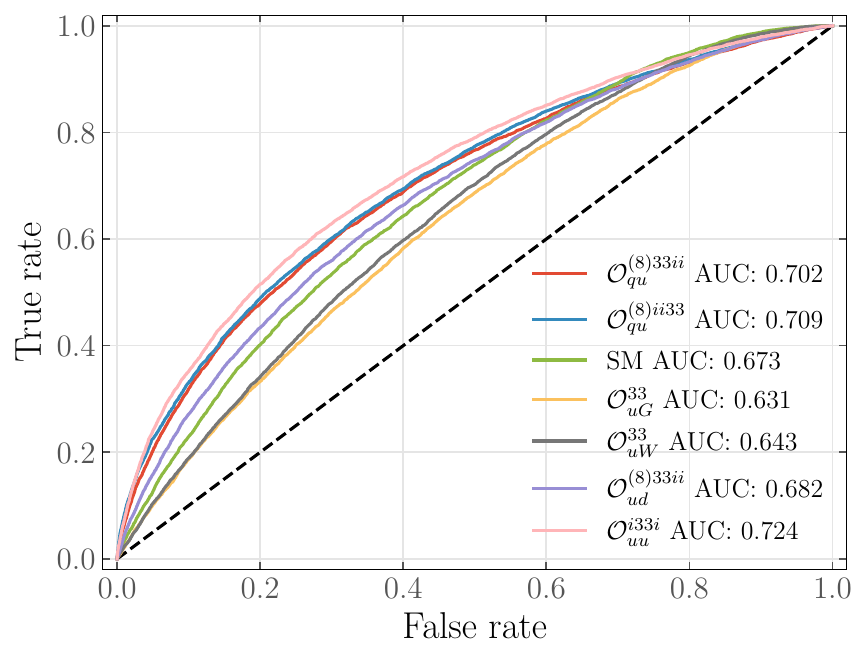}
	\caption{ROC curves for the scenario where multi-class classification is performed on thirteen SMEFT operators and the SM.}
	\label{fig:roc_all}
\end{figure*}

An alternative approach to formulating constraints is to directly employ the output of the GNN, i.e. using 2D histograms of the probabilities from the network (see, for example, the individual histograms from each contribution in~\cref{fig:2D_hist}), in place of the $p_T(b_1)$ distributions of~\cref{fig:ptb1_histos}. A $d$-dimensional classification can be converted into a $d-1$ dimensional probability histogram. This can act as a template for limit setting using the information that has condensed down the phenomenologically available information into the operator classification. Considering again ${\cal{O}}_{qu}^{(8)ii33}$ and ${\cal{O}}_{qq}^{(3)ii33}$, this is demonstrated in~\cref{fig:2D_hist}, where the three histograms can be used to construct a $\chi^2$, in the same way as the 1D $p_T(b_1)$ distributions, allowing the information of all three histograms to contribute. The resulting contour from this method is shown on the right plot of~\cref{fig:ML_Contours}. This method also improves the bounds on the WCs compared to the original $p_T(b_1)$ distribution analysis with no cuts on the probabilities required. This approach is feasible when we consider only a small (sub)set of the relevant interactions. Turning to the full $d-1$ dimensional histogram very quickly increases the statistical uncertainty. As can be seen from the qualitative similarity of the two approaches, a minimisation of
\begin{equation}
\text{P(SM)}=1 - \text{P(BSM)}
\end{equation}
appears to be adequate for multi-dimensional EFT analyses, particularly at luminosities below 3 ab$^{-1}$.

It should be noted that the one-dimensional $\text{P}(\text{BSM})$ histogram could be used to construct a $\chi^2$ as well, in order to obtain the contours on the $C_{qq}^{(3)ii33}-C_{qu}^{(8)ii33}$ plane. However, the sensitivity is limited compared to the other approaches along certain directions, as shown in \cref{fig:ML_Contours}, due to the loss of information in the projection of the two-dimensional output to a one-dimensional score. We therefore have not explored this approach further.

\subsection{Fit constraints with GNN selections}
\label{sec:full}
\begin{figure*}[!t]
	\centering
	\includegraphics[width=0.46\textwidth]{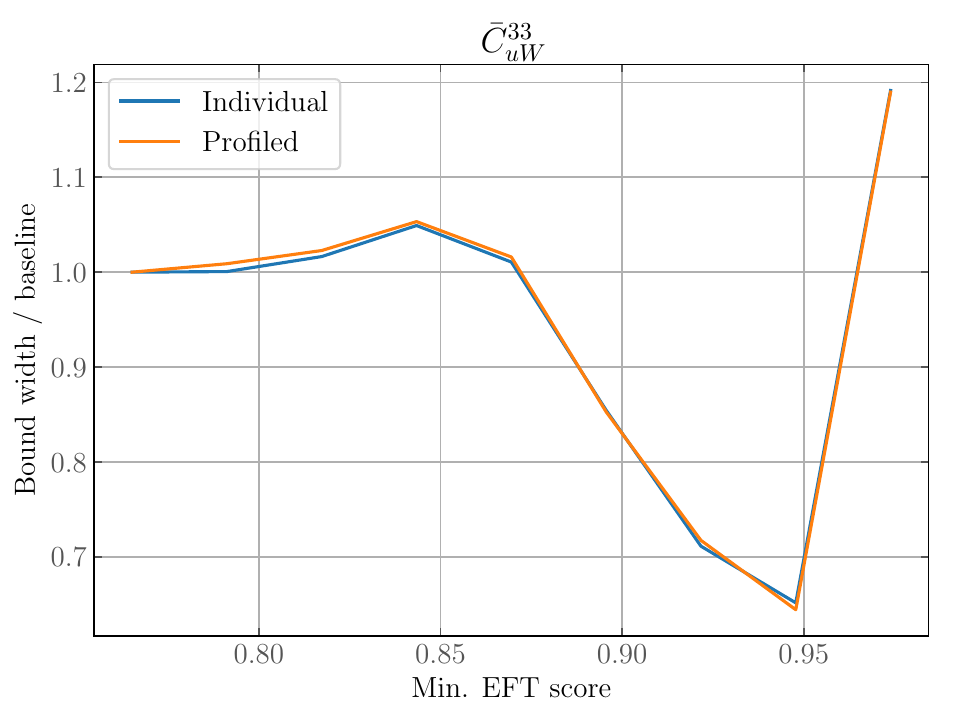}
	\hfill
	\includegraphics[width=0.46\textwidth]{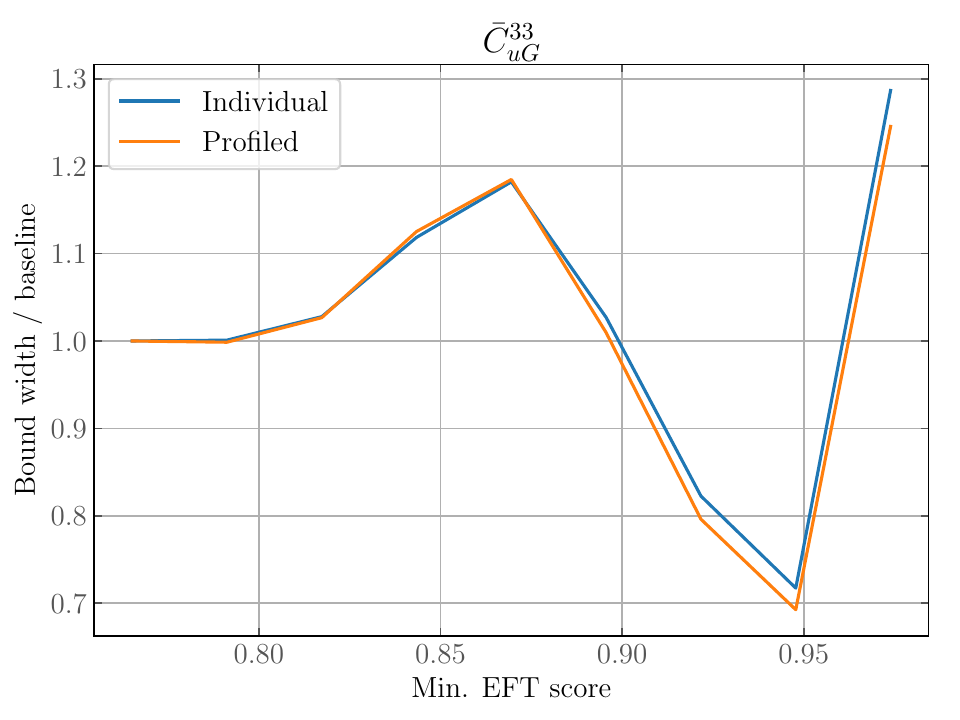}\\
	\includegraphics[width=0.46\textwidth]{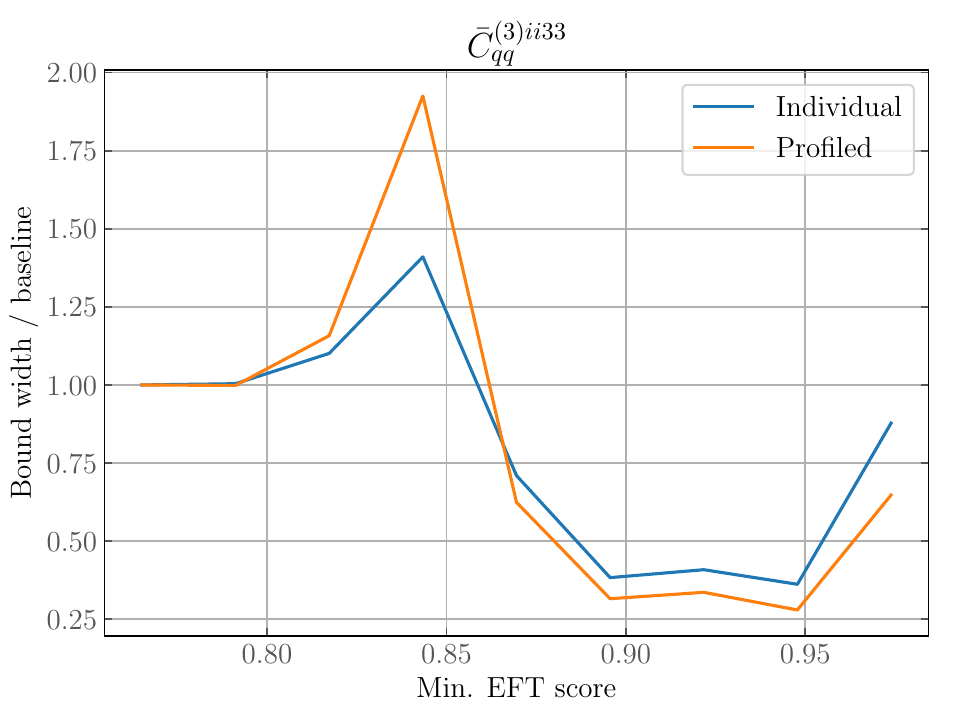}\hfill
	\parbox{0.46\textwidth}{\vspace{-5cm}
	\caption{Representative relative improvement (decrease in the 2$\sigma$ Wilson coefficient interval) over the individual (orange) and profiled (blue) operator constraints quoted in~\cref{tab:max_imp} by imposing cuts on the ML score. Bounds were obtained at an integrated luminosity of 3/ab.}\label{fig:indcut}}
	\end{figure*}

Extending the qualitative discussion of the previous section to the thirteen dimensional SMEFT parameter space, we show the Receiver Operator Characteristic (ROC) curves of the full classification in~\cref{fig:roc_all}. The ROC  curves are calculated with the generalised procedure discussed above. Again we see that the network\footnote{By optimizing the hyperparameters for this scenario we conclude that the architecture used for the two operators case continues to perform particularly well. Deeper networks do not significantly improve the performance and often suffer from longer training times and overtraining.} is capable of distinguishing operators adequately.

    \begin{table}[H]
        \centering
        \begin{tabular}{lRRRR}
            \toprule
            {} & \multicolumn{2}{c}{$2.3~\text{fb}^{-1}$} & \multicolumn{2}{c}{$3~\text{ab}^{-1}$} \\
            \cmidrule(lr){2-3}
            \cmidrule(lr){4-5}
            {}                             & \multicolumn{1}{c}{Individual} & \multicolumn{1}{c}{Profiled} & \multicolumn{1}{c}{Individual} & \multicolumn{1}{c}{Profiled} \\
            \midrule
            $\bar{C}_{G}$                  & (-0.0543, 0.0535)              & (-0.1785, 0.1776)            & (-0.0015, 0.0015)              & (-0.0047, 0.0047)            \\
            $\bar{C}_{\varphi q}^{(3) 33}$ & (-0.0317, 0.0326)              & (-0.0806, 0.0758)            & (-0.0009, 0.0009)				& (-0.0022, 0.0022) \\
            $\bar{C}_{uG}^{33}$            & (-0.0253, 0.0247)              & (-0.0622, 0.0655)            & (-0.0007, 0.0007)              & (-0.0017, 0.0017)            \\
            $\bar{C}_{uW}^{33}$            & (-0.0234, 0.0228)              & (-0.0544, 0.0580)            & (-0.0006, 0.0006)              & (-0.0015, 0.0016)            \\
            \midrule
			$\bar{C}_{qd}^{(8) 33ii}$      & (-0.1543, 0.1558)              & (-0.3789, 0.3698)            & (-0.0043, 0.0043)              & (-0.0104, 0.0104)            \\
            $\bar{C}_{qq}^{(1) i33i}$      & (-0.0202, 0.0204)              & (-0.0495, 0.0484)            & (-0.0006, 0.0006)              & (-0.0014, 0.0014)            \\
            $\bar{C}_{qq}^{(3) i33i}$      & (-0.0101, 0.0102)              & (-0.0247, 0.0241)            & (-0.0003, 0.0003)              & (-0.0007, 0.0007)            \\
            $\bar{C}_{qq}^{(3) ii33}$      & (-3.2964, 3.3259)              & \bdash                       & (-0.0917, 0.0917)              & (-0.3045, 0.3046)            \\
            $\bar{C}_{qu}^{(8) 33ii}$      & (-0.0867, 0.0875)              & (-0.2127, 0.2079)            & (-0.0024, 0.0024)              & (-0.0058, 0.0058)            \\
            $\bar{C}_{qu}^{(8) ii33}$      & (-0.0577, 0.0583)              & (-0.1416, 0.1383)            & (-0.0016, 0.0016)              & (-0.0039, 0.0039)            \\
            $\bar{C}_{ud}^{(8) 33ii}$      & (-0.1598, 0.1613)              & (-0.3923, 0.3824)            & (-0.0044, 0.0044)              & (-0.0107, 0.0107)            \\
            $\bar{C}_{uu}^{i33i}$          & (-0.0225, 0.0228)              & (-0.0553, 0.0540)            & (-0.0006, 0.0006)              & (-0.0015, 0.0015)            \\
            \midrule
            $\bar{C}_{lq}^{(3) ii33}$      & \bdash                         & \bdash                       & (-0.3289, 0.3288)              & (-1.8493, 1.8930)            \\
            \bottomrule
        \end{tabular}
        \caption{Baseline $2\sigma$ bounds for different luminosities.}
        \label{tab:baseline_bounds}
    \end{table}

Starting from the baseline sensitivity as quoted in~\cref{tab:baseline_bounds} (see also~\cref{sec:eft}), we first show how contributing operators are impacted
by imposing ML score cuts in~\cref{fig:roc_all}. Sizeable improvements can be obtained when the momentum enhancement is present (e.g. in case of $\bar{C}^{33}_{uG}$). Similarly, the graph network performs well in discriminating the non-resonant top decay contributions, e.g. in case  $\bar{C}^{33}_{uW}$. Improvements ranging between 5\% and 60\% are achievable in such instances~(see~\cref{tab:max_imp}), depending on the operators under consideration, however, always at stringent cuts on the ML score to achieve a generic BSM-sensitive selection (before losing statistical control for score cuts approaching unity). Representative operator improvements as a function of the ML score are given in~\cref{fig:indcut}. Operators showing a relatively small improvement are already under relatively good control via the inclusive rate and the baseline selection, which establishes good sensitivity to such non-SM interactions. In particular this holds for the $\bar{C}_G$ direction (which can be constrained in more adapted ways by exploiting multi-jet production~\cite{Dreiner:1991xi,Dixon:1993xd}). 

    \begin{table}[t]
        \centering
        \begin{tabular}{lRRRR}
            \toprule
            {} & \multicolumn{2}{c}{$2.3~\text{fb}^{-1}$} & \multicolumn{2}{c}{$3~\text{ab}^{-1}$} \\
            \cmidrule(lr){2-3}
            \cmidrule(lr){4-5}
            {}                             & \multicolumn{1}{c}{Individual} & \multicolumn{1}{c}{Profiled} & \multicolumn{1}{c}{Individual} & \multicolumn{1}{c}{Profiled} \\
            \midrule
            $\bar{C}_{G}$                  & 0.07 \%                          & 14.12   \%                      & 0.07   \%                         & 11.09      \%                   \\
            $\bar{C}_{\varphi q}^{(3) 33}$ & 33.74 \%                          & 34.19      \%                   & 33.73   \%                        & 33.48    \%                     \\
            $\bar{C}_{uG}^{33}$            & 28.29 \%                          & 32.18   \%                      & 28.28     \%                      & 30.74  \%                       \\
            $\bar{C}_{uW}^{33}$            & 34.86    \%                       & 35.35   \%                      & 34.85    \%                       & 35.53    \%                     \\
            \midrule
            $\bar{C}_{qd}^{(8) 33ii}$      & 4.71     \%                       & 4.68   \%                       & 4.71     \%                       & 4.76    \%                      \\
            $\bar{C}_{qq}^{(1) i33i}$      & 3.50     \%                       & 3.45   \%                       & 3.50     \%                       & 4.73    \%                      \\
            $\bar{C}_{qq}^{(3) i33i}$      & 4.35  \%                          & 4.28        \%                  & 4.35    \%                        & 5.00   \%                       \\
            $\bar{C}_{qq}^{(3) ii33}$      & 63.83   \%                        & \bdash                       & 63.83   \%                        & 71.91   \%                      \\
            $\bar{C}_{qu}^{(8) 33ii}$      & 3.45   \%                         & 3.51         \%                 & 3.45  \%                          & 3.48   \%                       \\
            $\bar{C}_{qu}^{(8) ii33}$      & 3.74   \%                         & 3.72         \%                 & 3.74   \%                         & 3.77  \%                        \\
            $\bar{C}_{ud}^{(8) 33ii}$      & 4.62   \%                         & 4.46         \%                 & 4.62   \%                         & 4.79  \%                        \\
            $\bar{C}_{uu}^{i33i}$          & 3.38    \%                        & 3.35         \%                 & 3.38    \%                        & 1.95   \%                       \\
            \midrule
            $\bar{C}_{lq}^{(3) ii33}$      & \bdash                         & \bdash                       & 10.57     \%                      & 35.52     \%                    \\
            \bottomrule
        \end{tabular}
        \caption{Maximum improvements in $2\sigma$ bounds via a cut on the ML score.}
        \label{tab:max_imp}
    \end{table}

Since individual constraints focus on one operator fixing the rest of the WCs to zero, it is common practice to profile over the rest of the WCs by determining their value such that the $\chi^2$ function is minimised. In the scenario where the analysis is particularly sensitive to the presence of any additional operator, a significant decrease in sensitivity will arise. We
calculate the improvement in the case of profiled WCs which, as shown in~\cref{fig:indcut}, remains similar to the individual WCs case. This is expected as the network selection removes background contributions but keeps new-physics effects. However, we note that the improvement on profiled bounds can be greater than on individual ones as in~\cref{fig:indcut}. This occurs when the cut on the EFT score selects a region where the impact on the bounds of a particular operator by the presence of additional ones is reduced, even though the robustness of one class against variations of others is not taken into account in our work.

\section{Summary and Outlook}
\label{sec:conc}
The absence of direct evidence for new physics beyond the Standard Model at the LHC is as surprising as it is challenging for particle physics.
Turning to effective field theory methods with the aim of fingerprinting new physics through the observation of modifications of expected SM correlations
in the plethora of LHC data is a well-motivated approach to experimentally challenge, and perhaps, overcome the current status quo. The multitude of ad hoc new physics interactions in the SMEFT approach demands tailored approaches to achieve the most sensitive limit setting. In this sense, limiting analyses to a handful of, albeit motivated, differential distributions is not beneficial for enhancing the sensitivity. Conversely, employing machine learning techniques that fingerprint and exploit correlations in data provides a highly adaptive avenue to enhance the overall sensitivity that can be achieved at the LHC but also other (future) collider experiments. 

In this work, we have focused employing on GNNs for EFT limit setting. GNNs are particularly motivated approaches for this purpose as they allow us to directly reflect the graph structure which is imposed by EFT interactions in the classification and eventual limit setting. We base our analysis on the semileptonic $t\bar t$ final states, as this is a motivated phenomenological arena for the presence of new interactions, but also because we face a critically large Wilson coefficient parameter space for multi-label classification. We find that large improvements of the sensitivity become achievable when correlations are not yet fully exploited in the inclusive base selection. This demonstrates that machine learning of multi-labelled collider data provides an excellent avenue towards improving the sensitivity of EFT-related measurements at colliders. We find that this improvement translates from individual to profiled bounds; our results also indicate a strategic approach to improve profiled constraints by tensioning operators against each other, which is not directly accessible by minimising the SM probability, but highlights the relative operator probabilities as another avenue for future investigations. Along these lines, we also note that optimisations of the ML score can be achieved via different weightings of the individual class probabilities. This way more model-specific (i.e. matched) interpretations of EFT constraints can be included to the machine learning stage, which should lead, in principle, to further sensitivity enhancements. 

We note that the results of our exploratory study presented here are based on a Monte Carlo analysis; the comparison of actual data with Monte Carlo predictions is affected by a range of theoretical and experimental uncertainties. While our results do not include such uncertainties, in principle it is possible to treat them via Generative Adversarial Neural Networks, e.g.~\cite{Louppe:2016ylz,Kansal:2021cqp}. Such an approach would discriminate between the different (labelled) hypotheses when the data is well-described by individual classes or superpositions of classes, effectively removing modelled uncertainty parameters from the classifier score. In general, this will lead to a decreased sensitivity compared to the idealised situation of the proof-of-principle analysis presented in this work. There are examples of such approaches to treat theoretical~\cite{Englert:2018cfo} and experimental~\cite{Bellagente:2019uyp} uncertainties. We leave modifications of the architecture presented in this paper along these lines for future work.

\section*{Acknowledgements}
This work is supported by the UK Science and Technology Facilities Council (STFC) under grant ST/T000945/1.
O.A. is funded by a STFC studentship under grant ST/V506692/1. 
S.B. is funded by a Scottish Data Intensive Science Triangle (ScotDIST) studentship under grant ST/P006809/1.
C.E. is supported by the IPPP Associateship Scheme and the Leverhulme Trust under grant RPG-2021-031.
P.S. is funded by a STFC studentship under grant ST/T506102/1.

\bibliographystyle{JHEP}
\bibliography{paper.bbl}

\end{document}